\documentclass[%
superscriptaddress,
preprint,
nofootinbib,
bibnotes,
 amsmath,amssymb,
 aps,
pre, longbibliography,
]{revtex4-1}

\usepackage{graphicx}
\usepackage{dcolumn}
\usepackage{bm}

\usepackage{gensymb}

\usepackage{orcidlink}

\newcommand{\orcidJuan}{0000-0003-3756-5016}
\newcommand{\orcidLeo}{0000-0002-9516-1346}

\begin{document}


\title{A synchrotron-like pumped ring resonator for water waves}

\author{\copyright Isis Vivanco}
\author{Alexander Egli}
\affiliation{Departamento de F\'isica, Universidad de Santiago de Chile,
Av. Victor Jara 3493, Estaci\'on Central, Santiago, Chile}

\author{Bruce Cartwright}
\affiliation{The University of Newcastle, Callaghan, New South Wales 2308,
Australia\\ Pacific Engineering Systems International, 277\textendash 279
Broadway, Glebe, New South Wales 2037, Australia}

\author{Juan F. Mar\'in\orcidlink{\orcidJuan}}
\email[]{j.marinm@utem.cl}
\affiliation{Departamento de Física, Facultad de Ciencias Naturales, Matemática y del Medio Ambiente, Universidad Tecnológica Metropolitana, Las Palmeras 3360, Ñuñoa 780-0003, Santiago, Chile.}

\author{Leonardo Gordillo\orcidlink{\orcidLeo}}
\email[]{leonardo.gordillo@usach.cl}
\affiliation{Departamento de F\'isica, Universidad de Santiago de Chile,
Av. Victor Jara 3493, Estaci\'on Central, Santiago, Chile}

\begin{abstract}
The wave-like behaviour of matter in quantum physics has spurred insightful analogies between the dynamics of particles and waves in classical systems. In this study, drawing inspiration from synchrotrons that resonate to accelerate ions along a closed path, we introduce the concept of a \emph{synchrowave}: a waveguide designed to generate and sustain travelling water waves within a closed annular channel. In analogy to unavoidable energy losses in conventional particle accelerators due to electromagnetic radiation and inelastic collisions, the system displays undesired water-wave dampening, which we address through the synchronised action of underwater wavemakers. Our analogies extend the resonance mechanisms of synchrotrons to generate gravity waves in closed waveguides efficiently. A proof-of-concept experiment at a laboratory scale demonstrates the unique capability of this technique to build up anomalously large travelling waves displaying a flat response in the long-wave limit. Besides quantifying the performance of wave generation, our findings offer a framework for both industrial and computational applications, opening up unexplored possibilities in hydraulics, coastal science and engineering. In a broader context, our experimental apparatus and methods highlight the versatility of a simple yet powerful concept: a closed-path continuous-energy-pumping scheme to effectively harvest prominent resonant responses within wave-supporting systems displaying weak dissipation.
\end{abstract}

\maketitle

\section{Introduction}

The concept of matter waves is a core idea introduced during the early stages of quantum physics. Some of the most intriguing properties of quantum systems useful in modern applications are rooted in wavelike phenomena. Quantum-like effects that are distinctive of wave dynamics, such as interference and tunnelling phenomena \cite{Martin1992, Papatryfonos2022}, confinement in quantum corrals \cite{Saenz2018}, quantised responses \cite{Perrard2014}, bound-state generation \cite{Hsu2016}, pilot-wave dynamics \cite{Bush2015}, quantisation of orbits \cite{Fort2010}, have been scrutinised in different fields of classical physics that support waves, ranging from electromagnetism to acoustics. However, there is no better playground to explore and visualise such analogies than fluid mechanics \cite{Couder2005, Couder2006, Eddi2009, Eddi2012, Labousse2016}. This stands even though energy dissipation in fluids is orders of magnitude larger than at a quantum scale. Indeed, water-wave-decay inhibition \cite{Cobelli2009, Monsalve2019} is possible and has been achieved by using experimental configurations that compensate energy losses through moving boundaries \cite{Cobelli2009, Vivanco2021}, leading to inviscid-like water-wave behaviour. Similar mechanisms have been helpful in studying energy transmission in tsunami events \cite{Jamin2015},  internal waves in stratified fluids induced by tides\cite{Maurer2017}, and nature-inspired locomotion strategies in viscous flows \cite{Pandey2023}. 

Energy compensation in weakly dissipative systems often implies the existence of resonances, i.e. abnormal amplifications of the system response at given frequencies. Under certain conditions, a resonance can be engineered to devise valuable applications. For instance, synchrotrons are particle accelerators operating under the resonance and synchronisation principles. Their use as a radiation source is widely spread for applications in fundamental research, applied physics, and the medical industry\cite{Nielsen2011}. In a synchrotron, ions emitted from a particle source accelerate while restrained to travel along a closed path at constant speed. The curved trajectory is achieved through the action of a static magnetic field. However, ions on a curved path lose energy through electromagnetic radiation. Ions also suffer inelastic scattering with other particles within the ring, which are present even under ultra-high vacuum conditions \cite{Baglin2024}. Such energy losses are compensated through the action of an alternating electric field applied at every segment, or building block, of the closed path. When properly synchronised with the electrical field in each building block, the synchrotron becomes resonant, and the ions describe sustained propagation in a closed path at constant speed under the action of the Lorentz force.

Inspired by the motion of charged particles in a synchrotron, here we propose and study a synchrotron of water waves, or \emph{synchrowave}, a device that builds up gravity waves on a channel and sustains its propagation along a closed circular path at a constant speed. Inspired by the segmented alternating electric fields that drive the particles in conventional synchrotrons, a synchrowave sustains travelling waves by pumping energy at discrete building blocks through synchronous periodic actuation of servomotors under a soft bottom. Such periodic motion uses the recently proposed wave generation technique coined as pedal wavemaking \cite{Vivanco2021}. We study the linear response of surface gravity waves by analysing the hydrodynamic equations for viscous fluids under appropriate boundary conditions. We show that the synchrowave displays resonance, wave synchronisation, and filtering-like behaviour as a function of the wavelength, frequency, and viscosity. To round off, we designed and set up a tabletop synchrowave with 64 building blocks to test our hypotheses. We demonstrate that under resonant conditions, a small amplitude motion of the servomotors sustains large amplitude waves at the surface. Our experimental results agree quantitatively with our theoretical findings and match Smoothed-Particle Hydrodynamics numerical simulations as well.

\section*{The synchrotron-like water waveguide}

\begin{figure}
    \centering
    \includegraphics[width=\linewidth]{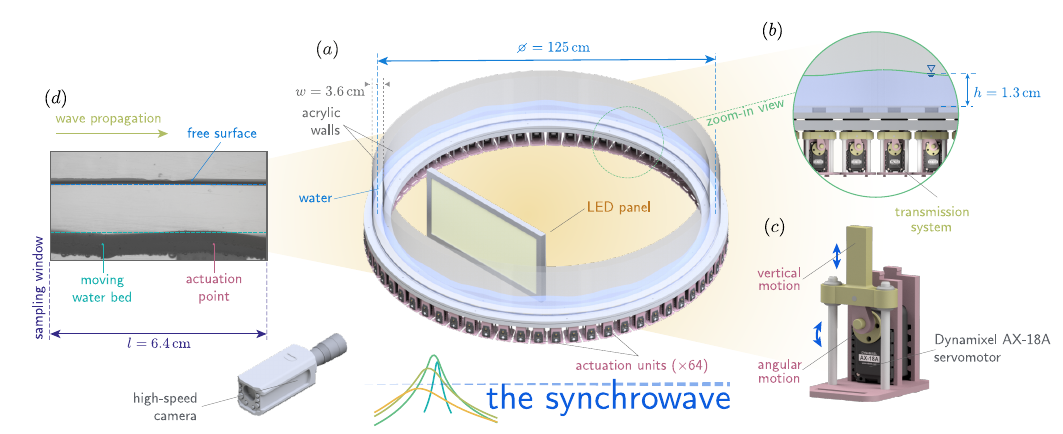}
    \caption{\textbf{The Synchrowave: a synchrotron water waveguide machine.} \textbf{(a)} Three-dimensional scheme of the synchrowave. The annular waveguide channel for gravity waves consists of two annular walls and a soft bed on a set of 64 servomotors uniformly spanned along the bottom for multipoint action. The free-surface motion is recorded using a high-speed camera and an LED panel. \textbf{(b)} The zoom-in view highlights the servomotors' arrangement below the soft bed. \textbf{(c)} A zoom-in on the base of each unit shows every component involved in motion transmission, which allows independent control. \textbf{(d)} Snapshot of a sampling window on the experimental system under the action of the moving water bed.}
    \label{Fig:01}
\end{figure}

Our proof of concept of a synchrowave designed to sustain travelling water waves with a programmed wavelength is shown in Fig.~\ref{Fig:01}(a). The water is confined in a 125-cm-diameter annular waveguide or channel, 3.6-cm wide, made of laser-cut acrylic cylindrical walls and 3d-printed synthetic-polymer (PLA) and plastic-ABS parts, as shown in Fig.~\ref{Fig:01}(b). A soft elastomer cast seal bed on the base is pressed to ensure a waterproof and snug fit, as shown in Fig.~\ref{Fig:01}(b).  

The synchrowave is assembled from 64 building blocks, each including a servomotor fixed to the soft-bed section of the channel. Each servomotor has an individual transmission system with a connecting rod and a cam for converting servomotor rotational motion to a vertical push. The transmission system also comprises linear bearings and a piston, as shown in Fig.~\ref{Fig:01}(c). The purpose of the servomotor array is to pump energy constructively to the waves as they travel, thus compensating for their losses. Servomotors are wired to power supplies and connected to a central microcontroller that establishes a daisy-chain communication protocol for independent control. We placed the channel between a high-speed camera and an LED panel, as shown in Fig.~\ref{Fig:01}(a), to record the dynamics of the free surface in a window (see Fig.~\ref{Fig:01}(a,d)).

\begin{figure}
    \centering
    \includegraphics[width=\linewidth]{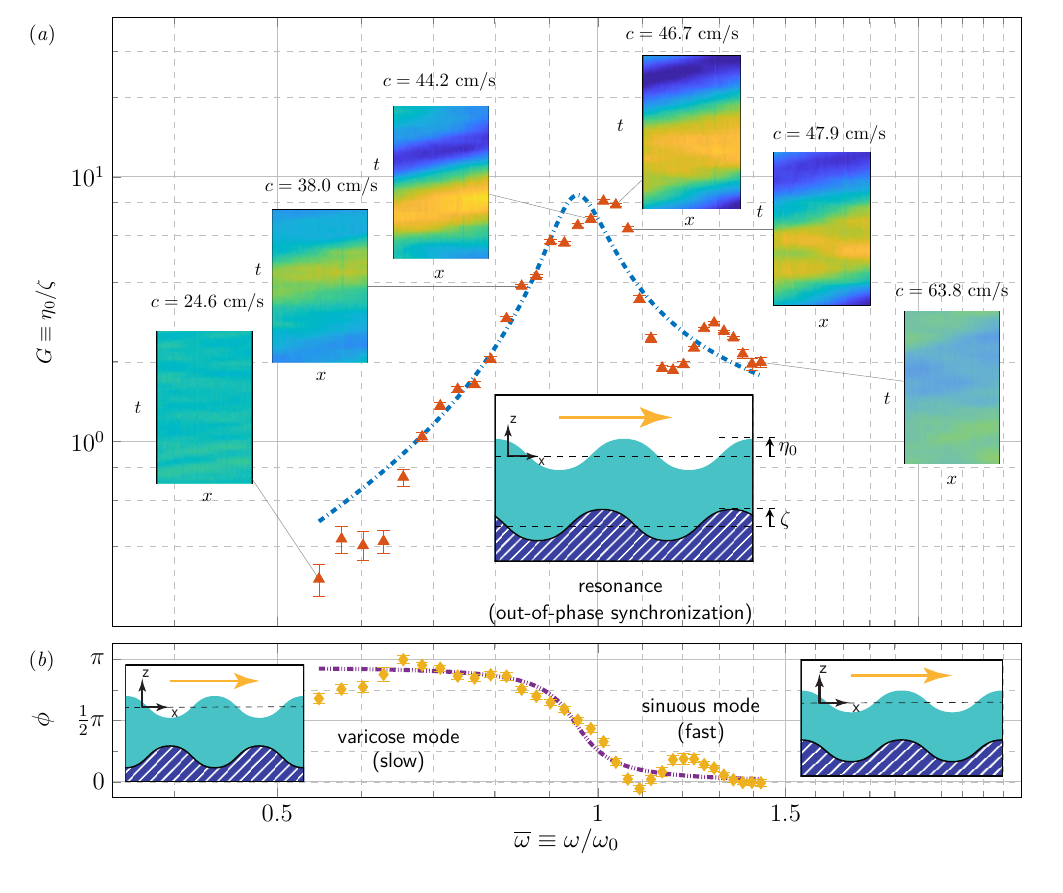}
    \caption{\textbf{Resonant response of gravity waves in a synchrotron water waveguide:} Experimental results in a water channel of depth $h=1.3$~cm for a bottom actuation of wavelength $\lambda/R = \pi/8 $  ($\lambda \simeq 49.1$~cm) at an amplitude $a=1$~mm. Servomotors are synchronised to oscillate at different frequencies, keeping the traveling-phase scheme. \textbf{(a)} Gain (wave-to-bottom displacement ratio) of the system as a function of the dimensionless forcing frequency. Theoretical prediction (dash-dotted blue line) and experimental measurements (red triangles) exhibit a resonance peak. Spatiotemporal diagrams at different forcing frequencies are shown in the insets, evidencing the emergence of travelling gravity waves on the free surface around the resonant condition. The speed $c$ of the wave is shown at the top of each snapshot.  \textbf{(b)} Phase between the wave and the bottom displacement as a function of the driving frequency of the servomotors.}
    \label{Fig:02}
\end{figure}

For each experimental run, we start with the fluid at rest and slowly increase the servomotors' oscillatory motion amplitude at both fixed input frequency and wavelength. Once the fluid's surface wave amplitude has reached a stable value, we use a boundary-detection image algorithm to measure the fluid displacement, and hence, the wave amplitude $\eta_0$. The details of the experimental protocol and image processing are described in Methods. Figure~\ref{Fig:02} summarises our experimental characterisation of the synchrowave response to the bottom periodic motion using an eight-servomotor wavelength, corresponding to $\lambda\approx 49.1$~cm. 
Figure~\ref{Fig:02}(a) shows the system gain in amplitude $G$, defined as the ratio of the wave amplitude $\eta_0$ to the bottom displacement $\zeta$, as a function of the dimensionless input frequency $\overline \omega \equiv \omega/\omega_0$, i.e. the ratio of the forcing frequency $\omega$ to the natural frequency $\omega_0$, which depends on $\lambda$ via the dispersion relation of water waves on uniform depth, $\omega_0(k)=\sqrt{gk\tanh{kh}}$, Here, $k=2\pi/\lambda$ and $h=1.3$~cm is the water depth. The data shows that the travelling gravity wave gradually builds up as the frequency of servomotors increases from $f=0.5$ Hz up to $f=0.875$ Hz, where the amplitude of the gravity wave reaches its overall maximum value. The maximal gain is placed at $\omega=\omega_{\mathrm{res}}$, close to the natural frequency condition $\overline \omega=1$. 

Synchrowave-generated waves have the same frequency of oscillations as the bottom motion. Since the amplitude of the servomotors' oscillations $a$ is small compared to the wavelength ($\lambda/a\simeq491$), the system is linear and responds at the same input frequency. Yet,  different synchronisation regimes exist, as evidenced by the phase shift between the servomotors' motion and the generated gravity wave. Figure~\ref{Fig:02}(b) shows the phase shift of the wave as a function of the servomotors' frequency. We observe the emergence of a slow mode for frequencies far below the resonance condition, $\omega\ll\omega_{\mathrm{res}}$, where the free surface and the soft bottom slowly oscillate in anti-phase, thus generating a varicose fluid layer (see left inset in Fig.~\ref{Fig:02}(b)). When increasing the frequency, the system approaches the resonance condition. At resonance, $\omega=\omega_{\mathrm{res}}$, the surface-wave and servomotor displacement become out-of-phase (see bottom inset in Fig.~\ref{Fig:02}(a)). Far above the resonance condition, $\omega\gg\omega_{\mathrm{res}}$, we observed the emergence of a fast mode, in which the free surface and the servomotors oscillate in phase. In this latter case, the synchronisation establishes a sinuous fluid layer in which the water wave replicates the bottom displacement (see right inset in Fig.~\ref{Fig:02}(b)).



After having depicted our experiment and summarised its main outcomes, we draw a parallel between conventional synchrotrons and synchrowaves, emphasizing their operating mechanisms’ resemblance and differences (see Table~\ref{Table:01}). Further details are provided in Methods.   

\scalebox{0.4}{\includegraphics[]{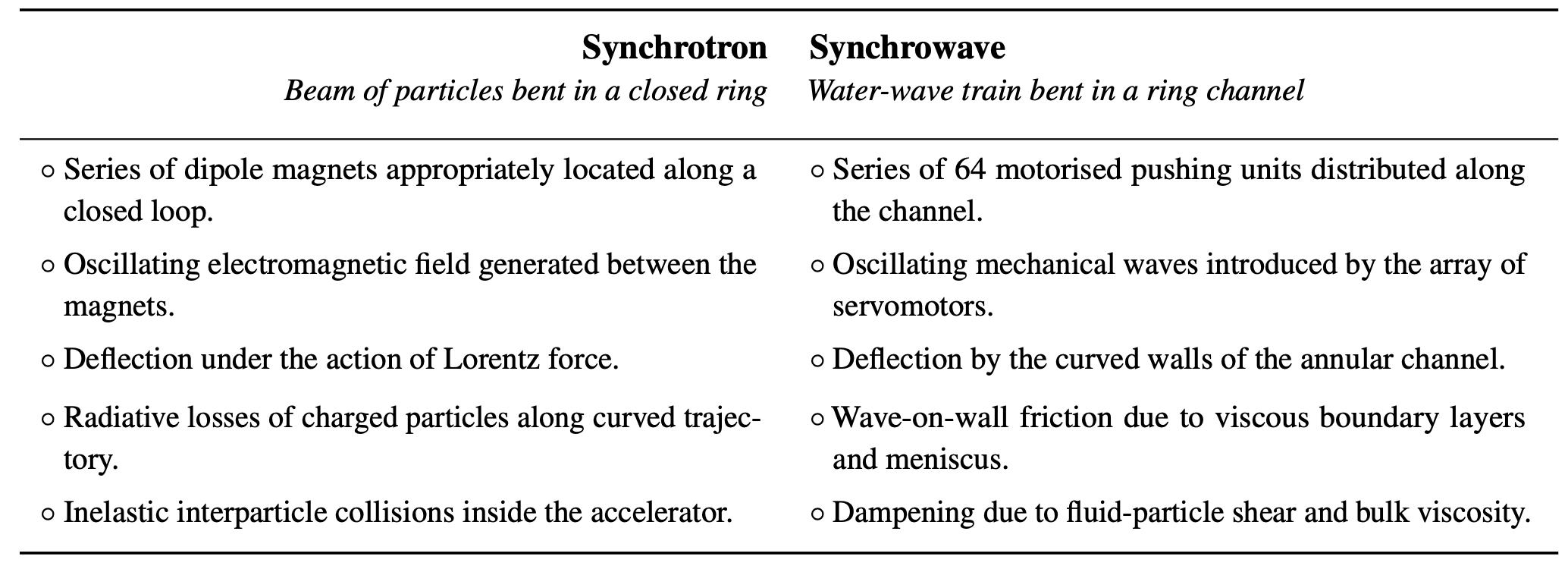}}
{\small{}{\textbf{TABLE 1: Analogies between synchrotrons and synchrowaves:} Parallel
between the constituent parts and the physical processes in charged-particle
synchrotrons and water-wave synchrowaves.}
\label{Table:01} }{\footnotesize\par}

\section*{Synchrowave theoretical modelling}
\label{Sec:Resonance}

To provide further insights into the dynamic response of synchrowaves under general synchronisation schemes, we model the synchrowave as a fluid layer of uniform depth $h$ in the $x-z$ plane and kinematic viscosity $\nu$, under the action of gravity, $\mathbf{g}=-g\mathbf{k}$. The bed on which fluid lies is at $z=-h$ and is deformable, i.e., it is allowed to displace and push the liquid above. The model neglects the effect of the curvature of the waveguide of the synchrowave, as it is negligible compared to the wavenumber of the water waves.  A description and quantitative justification for the accounted simplification is provided in Methods. 

\begin{figure*}
\begin{center}
\includegraphics[width=\linewidth]{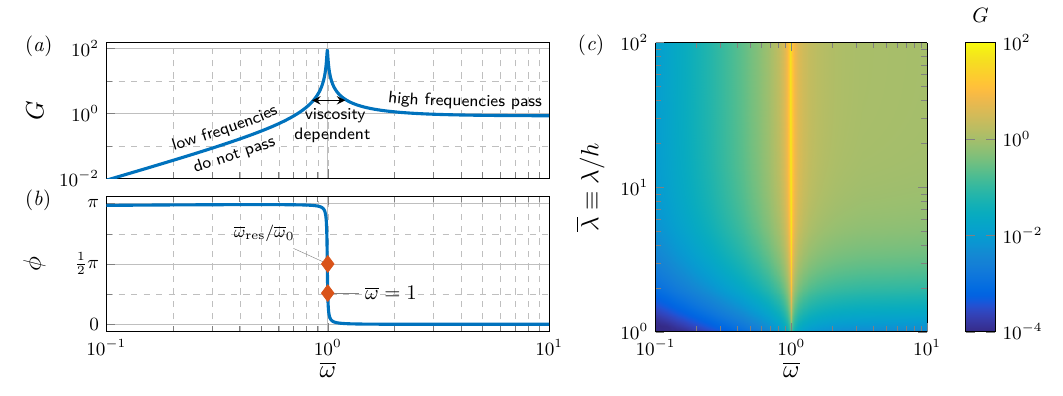}
\end{center}
 \caption{\textbf{Theoretical frequency-response curves:} Prediction for a $40\,\mathrm{m}$-deep channel filled with a fluid of viscosity $\nu=1800\,\mathrm{cSt}$ ($\overline\Theta=4.55\times10^{-4}$). \textbf{\textit{(\textit{a})}}
 The gain of the synchrowave for $\overline\lambda=10$, exhibiting a resonant peak near the frequency $\omega_0=\sqrt{gk\tanh{kh}}$. The system behaves as a filter of low frequencies. \textbf{(\textit{b})} The phase $\phi=-\arg(Y)$, revealing
 synchronisation at frequency ranges above the resonance condition (indicated by diamonds). \textbf{(\textit{c})} Gain of the system as a function of the wavelength and the frequency of the bottom motion. \label{Fig:03}}
\end{figure*}

A vertical travelling-wave displacement introduced by the servomotors at the base can thus be described by the complex field $z_b$, given by
\begin{equation}
z_b(x,t)=\zeta\exp[\mathrm{i}\left(kx-\omega t\right)],\label{Eq:InputVertical}
\end{equation}
where $\zeta$ is the complex amplitude of the forced vertical motion, and the phase propagates in the positive $x$-direction (hereon the streamwise direction), with prescribed wavenumber $k$ and angular frequency $\omega$. We use the dimensionless wavenumber $\overline{k}\equiv kh$ and frequency $\overline{\omega}\equiv\omega/\omega_0$. Likewise, the amplitude of the surface wave, the vertical motion of the servomotors at the bottom, and the space coordinates in dimensionless form are $\overline\eta_0=\eta_0/h$, $\overline\zeta=\zeta/h$, $\overline{z}=z/h$, and $\overline{x}=x/h$, respectively. Solving the linearised system (see Methods), we obtain the gain $G$, i.e. the ratio of the free-surface-wave complex amplitude $\overline\eta_{0}$ to bed vertical-oscillation amplitude $\overline\zeta$, in terms of the parameters of the system, given by

\begin{equation}
\label{YAmplitude}
 G(\overline\omega, \overline k)=\frac{\left(1+\mathrm{i}\overline\Theta\right)\cosh\alpha\overline{k}-\mathrm{i}\overline\Theta\cosh\overline{k}}{\cosh\alpha\overline{k}\cosh\overline{k}\left(1-\frac{1}{\overline\omega^{2}}\right)\left(1-\frac{\mbox{tanhc}\,\alpha \overline{k}}{\mbox{tanhc}\,\overline{k}}\right)+\frac{\mbox{sinhc}\,\alpha \overline{k}}{\mbox{sinhc}\,\overline{k}}
+\overline{k}^{2}\left[\mbox{sinhc}^2\left(\frac{\overline{k}(\alpha +1)}{2}\right)+\mbox{sinhc}^2\left(\frac{\overline{k}(\alpha -1)}{2}\right)\right]}.
\end{equation}
Here, $\overline \Theta \equiv 2\nu/\sqrt{gh^3}$ is the dimensionless viscosity of the fluid, and $\alpha \equiv \sqrt{1-i\omega/\nu k^2}$ is a dimensionless complex parameter. The functions $\mbox{sinhc}(x)=(\sinh x)/x$ and $\mbox{tanhc}(x)=(\tanh x)/x$ are the cardinal hyperbolic sine and tangent, respectively \cite{Sanchez2012}. Figure~\ref{Fig:03}(\textit{a}) shows that the gain derived from equation~\eqref{YAmplitude}, exhibits a resonance peak for $\overline\omega=\overline\omega_{\footnotesize\mbox{res}}\simeq1$, i.e, for $\omega\simeq\omega_0$. At such precise synchronous conditions, the motion is optimally transferred from the servomotors to the surface of the fluid, generating the largest amplitude wave able to circulate indefinitely in the channel as long as the action of the servomotors is sustained. This mechanism has a complete analogue in conventional synchrotrons, where beam losses are compensated through alternating electric fields synchronously applied with a precise (resonant) frequency through radio-frequency cavities. These points of input of energy must be conveniently placed along the circular path of the curved beam. Indeed, the spatial separation between such cavities and the phase difference between their emitted fields induce an effective space dependence and wavelength of the alternating electric field along the loop. The analogy between this mechanism in conventional synchrotrons and the "field" generated by the motion of our array of servomotors thus becomes evident.

Figure~\ref{Fig:03}(\textit{a}) also shows that the synchrowave filters slow frequencies as there is a persistent gain decay under $\overline\omega_{\footnotesize\mbox{res}}$, while in the fast frequency spectrum limit, the gain reaches a plateau. The system's temporal asymptotic response, far from the resonance, is, hence, that of a high-pass filter, a behaviour reported for impulsively forced water waves in the context of tsunami generation \cite{Kajiura1963, Hammack1973, Jamin2015}. Likewise, the theoretical model can predict the synchronisation modes as a function of frequency. The Bode phase diagram shown in Fig.~\ref{Fig:03}(\textit{b}) shows the phase $\phi=-\arg(\overline\eta_0)$ of the gravity wave as a function of the input frequency $\overline\omega$. It shows clearly that the wave synchronises in two different modes: Far below the resonant frequency, the fluid layer propagates through the waveguide in a varicose mode, while far above, it follows the bottom in a sinuous mode.  

To contrast our theoretical results, we appended in Fig.~\ref{Fig:01} our predictions for the gain and the phase (dash-dotted lines), for easy comparison with the synchrowave experimental data. Given that our theoretical model does not consider viscous effects due to the friction with the lateral walls, we used fluid viscosity as a parameter to be fitted, keeping the remaining ones matched with our experimental conditions. The agreement between theory and experiments is remarkable in both the gain and the phase response.

\begin{figure*}
\begin{center}
\includegraphics[width=\linewidth]{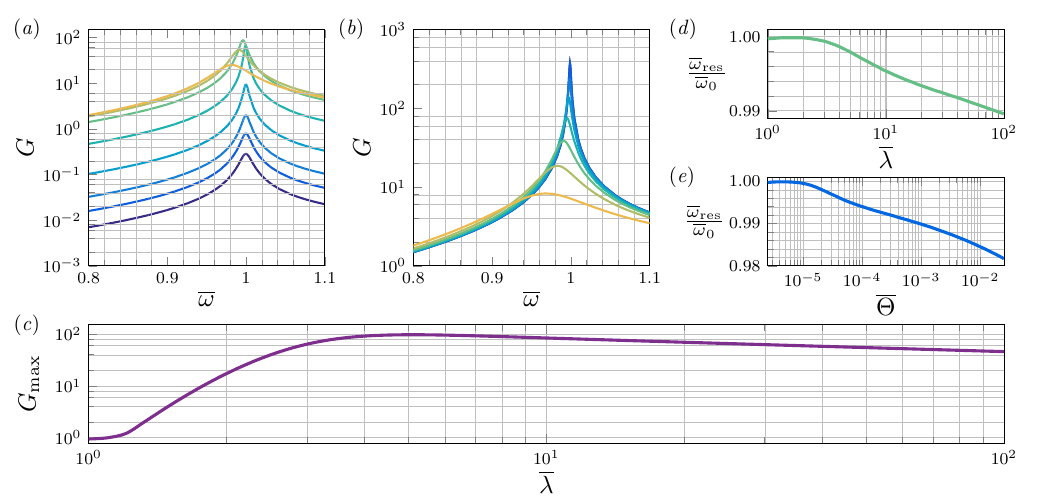}
\end{center}
 \caption{\textbf{Performance of the synchrowave:} Synchrowave response as a function of viscosity and wavelength following our theoretical modelling. Gain as a function of the dimensionless frequency for different values of: \textbf{(\textit{a})} the dimensionless wavelength $\overline \lambda = [1,20]$ at fixed dimensionless viscosity $\overline\Theta=4.5\times10^{-4}$; and \textbf{(\textit{b})}, the dimensionless viscosity $\overline \Theta = 4.5 \times [10^{-7},10^{-3}]$ at fixed wavelength $\overline\lambda=10$
 \textbf{(\textit{c})} Maximum gain in the synchrowave, $G_{\max}$, vs. wavelength for $\overline \Theta=7.15 \times 10^{-6}$.  \textbf{(\textit{d})} Frequency of the resonant mode as a function of the dimensionless wavelength $\overline \lambda$ and \textbf{(\textit{e})} the dimensionless viscosity $\overline \Theta$. \label{Fig:04}}
\end{figure*}

The model provides a mechanical characterisation of the system under general conditions. Figure~\ref{Fig:03}(\textit{c}) summarises the synchrowave gain as a function of both the input frequency and wavelength, evidencing how synchrowaves filter short waves. For a long wavelength, the gain remains almost constant [see top left plateau in Fig.~\ref{Fig:03}(\textit{c})] reaching saturation around $\overline\lambda\simeq10$, and plumbing when the wavelength decreases further. Thus for a random distribution of wavelengths and frequencies far enough from the resonance, only those components with high frequencies and long wavelengths will build up. This property may find relevant applications in the recreation of oceanic waves in engineered tanks, given that ocean waves due to storms, earthquakes, and tides are long \cite{Munk1950}.

Besides displaying the high-pass filter and resonant behaviour for any forcing wavelength, the synchrowave has an overall optimal wavelength, as revealed by Figure~\ref{Fig:04}(a). Figure~\ref{Fig:04}(b) shows the intuitive but still noteworthy feature that the resonance peak becomes prominent and sharper as the viscosity decreases. At fixed viscosity, the response also displays a maximum gain $G_{\max}$ for each input frequency. $G_{\max}$ displays an overall high-gain low-pass filter behaviour when plotted against $\lambda$, as depicted in Fig.~\ref{Fig:04}(\textit{c}). Likewise, Fig.~\ref{Fig:04}(\textit{d}) shows that the resonance frequency $\overline\omega_{\footnotesize\mbox{res}}\sim1$ in a two-orders-of-magnitude-span wavelength interval.  Finally, we also found in our theoretical expressions that the properties of the waves are very robust to viscosity effects. Figure~\ref{Fig:04}(\textit{e}) shows that $\overline\omega_{\footnotesize\mbox{res}}\sim1$ inside a wide range of small viscosity values, $\overline\Theta\in[10^{-6},\, 10^{-2}]$ for an arbitrary wavelength $\overline\lambda=10$. 

\section*{Numerical simulations}

To test our results, we performed numerical simulations on a fluid layer with periodic boundary conditions along the streamwise direction using the \emph{Smoothed Particle Hydrodynamics} (SPH) method \cite{Lucy1977, Gingold1977}. All our simulations consider a fluid of density $\rho_0=1\,\hbox{g/cm}^3$ in a domain of depth $h=1\,\hbox{cm}$ and length $L=20\,\hbox{cm}$, under the action of a moving bottom. The details of the numerical methods are given in Methods. We study the SPH system response to servomotors in an input frequency span $\overline\omega\in[0.1,\,10]$ and fixed $\overline\lambda=10$.

\begin{figure}
    \centering
    \includegraphics[width=\linewidth]{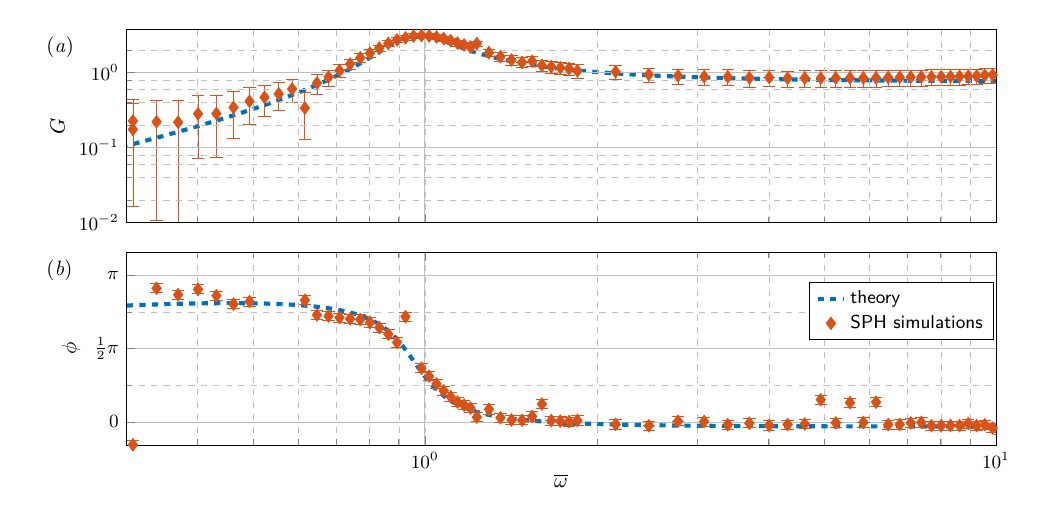}
    \caption{\textbf{Gain and phase of the synchrotron water waveguide:} Results for $\overline{\lambda}=10$. Comparison between SPH simulations (red diamonds) and analytical results using equation~\eqref{YAmplitude} (dashed blue line). Theoretical curves corresponds to $\overline{\Theta}=0.0958$ ($\nu=1.5\,\mbox{cSt}$).}
    \label{Fig:06}
\end{figure}

Figure \ref{Fig:06} presents the results from SPH simulations. In Fig.~\ref{Fig:06}(\textit{a}), we confirm that the synchrotron water waveguide exhibits typical resonance features, where the gain reaches the value $G_{\max}=3.1143$ at $\overline{\omega}=\overline{\omega}_{\hbox{res}}$. Moreover, SPH simulations confirm the high-pass filtering behaviour of the system: the gain strongly decays for decreasing values of $\overline{\omega}$ below $\overline{\omega}_{\hbox{res}}$. In contrast, if the frequency is increased above the resonance, the gain asymptotically decays towards unity. Given that our SPH simulations have been purposedly made highly viscous, the resonant peak is relatively broad, and the resonant frequency is slightly below the value corresponding to the inviscid case, $\overline{\omega}=1$. Such behaviour was previously observed in Fig.~\ref{Fig:04}(\textit{e}) in our model. In Fig.~\ref{Fig:06}(\textit{a}), we show with a dashed blue line the predicted curve following equation~\eqref{YAmplitude} for the gain corresponding to $\overline{\Theta}=0.0958$, which is in remarkable agreement with SPH simulations.

Finally, Fig.~\ref{Fig:06}(\textit{b}) shows the phase as a function of frequency obtained from SPH simulations, along with the corresponding theoretical curve for $\overline{\Theta}=0.0958$. We introduced a shift $\Delta\phi=0.21$ in the theoretical curve of the phase to consider numerical effects due to the artificially introduced compressibility of fluid particles in our SPH formulation \cite{Hughes2010}. We observe that the resonance condition occurs slightly below $\overline{\omega}=1$, consistent with our previous observations. SPH simulations confirm that the frequency of surface waves synchronises with the servomotors at fast frequencies.

\section*{Conclusions}
\label{Sec:Conclusions}

In this work, we emulated the resonance phenomenon that occurs in a synchrotron of charged particles into a new synchrotron of gravity water waves. Instead of a system where an oscillating electromagnetic field is induced at discrete magnetic cavities distributed along a circular path for accelerating charged particles, we designed a ring-shaped water channel made of discrete mechanical actuators, or building blocks, each one including a servomotor that pushes water from the bottom of the channel to excite and sustain gravity-wave trains at a constant speed. We have built a lab-sized synchrowave using 64 servomotors uniformly distributed along the bottom of a 125-cm-diameter annular waveguide. We analysed the response of travelling waves as a function of the synchronisation properties of the servomotors. We studied and characterised the resonance of gravity graves generated at the free surface of the viscous fluid using experimental data measured in the synchrowave and in simplified theoretical and numerical models. For our theoretical study, we analysed the linear Navier-Stokes equations in an infinite two-dimensional fluid layer of finite depth, providing a closed-form expression for the system gain as a function of fluid and bottom-motion parameters. The system exhibits a resonance peak and behaves like a high-pass filter for frequencies and a long-pass filter for wavelength. The resulting resonant peak can be sharp for small viscosity and display remarkable amplification factors. Our theoretical model accurately captures the observed experimental frequency response of the synchrowave and shows that one can always tune the frequency at which the system resonates for a given wavelength. Our main result is that synchrowaves can generate large amplitude long waves efficiently, with orders-of-magnitude amplifications for ordinary conditions.

Fluid parameters naturally affect the resonance: Increasing viscosity value decreases the system gain, widens the resonance peak and introduces a slight resonant-frequency shift. Accordingly, we have also shown an optimal wavelength for a given combination of viscosity and frequency at which the gain attains the maximum possible value. Thus, our wave-generation setup is efficient and can accurately mimic wave dynamics. Since our results can be used to design water waves with an on-demand wavelength, even in viscous fluids in narrow water channels, we test our hypotheses numerically using a basic formulation of SPH, which can be a highly viscous method without modern correction techniques. We performed a complete numerical characterisation of the resonance, showing excellent agreement with the theoretical predictions even at large viscosity.

Our theoretical modelling within the framework of dimensionless variables can be scaled to describe physical systems significantly larger than our tabletop synchrowave. A scaled-up water synchrowave should be able to display even sharper resonances with colossal gains. The results of this work suggest promising applications in the study of oceanic phenomena involving long gravity waves at the surface of the sea, such as ordinary tides, trans-tidal waves, and other waves generated by storms and earthquakes. In addition, our results may shed light on the open question of how relatively small-amplitude perturbations generate large-amplitude surface wave events, such as super-tsunamis appearing after small-intensity earthquakes and rogue waves appearing without warning in otherwise benign conditions.

It is still shocking how the simple, even obvious, scheme of pumping bits of energy along a closed path to build up amplitude after every turn can end in huge outcomes, especially when the energy losses are weak. The synchrowave, as well as the synchrotron, exploit this simple idea. One could wonder if such a scheme exists unprompted in nature. The Drake Passage, which separates South America's southern tip and Antarctica's northern tip, creates the only water belt across the globe with no relevant landmasses within. In this ocean region extending from the latitudes between parallels 50$^\circ$ and 60$^\circ$ south, circular unidirectional currents and tides find enough room to create free-surface resonant waves. For centuries, sailors have known this region for being almost impossible to sail due to their unexpected surges, rogue waves and unpredictable weather \cite{Shackleton1919}. A 19th-century sailor saying may provide an emotional yet accurate account of drifting on a planetary-scale synchrowave: \emph{Beyond 40 degrees south, there is no law. Beyond 50 degrees south, there is no God}.

\section{Methods}
\label{sec:Methods}

{\small 

\section*{The synchrowave: a synchrotron-based design}

Our experimental development is inspired by a fluid mechanics analogue of synchrotrons of charged particles, already summarised in Table~\ref{Table:01}.
In conventional synchrotrons, beams of accelerated particles are bent between dipole magnets. The trajectory of particles is deflected under the Lorentz force, and the storage ring is achieved by arranging a series of such bending magnets with appropriate separation around a closed loop. In our synchrowave, the curved walls of the annular channel naturally deflect the generated waves, closing their path after a turn in a waveguide. Curvature in synchrotrons and synchrowave introduces an effective loss. Indeed, the synchrotron's charged particles under the Lorentz force emit radiation as their direction of propagation changes, thus leading to a slow energy decay. In complete analogy, the walls on our synchrotron water waveguide also lead to wave losses in gravity waves due to viscous boundary layers and meniscus, leading to effective friction between sloshing waves and the curved walls.
Moreover, particles in conventional synchrotrons undergo unavoidable inelastic collisions with other particles in the accelerator, even under the best high-vacuum conditions achieved to date \cite{Baglin2024}. The analogous loss in our synchrowave machine is dampening due to viscosity in the bulk, which accounts for internal inelastic processes within the fluid due to adjacent-fluid layers' relative motion. The 64 building blocks distributed along the channel with servomotors compensate for the loss mechanisms. The oscillating mechanical waves introduced by the array of servomotors are equivalent to the oscillating electromagnetic fields generated between the magnets conveniently distributed in conventional synchrotrons. The critical point of synchrotrons and our synchrowave is the choice of a synchronisation scheme for the oscillating field to enable an optimal transfer of input energy to sustain the circular orbits at resonance. 

\section*{Experimental development}

The synchrowave is a $3.6$-cm-wide channel along a $125$-cm-diameter circumference enclosed by two  $30.0$-cm-high concentric walls. All pieces and parts of the synchrowave were designed using \textsc{SolidWorks}\copyright, and manufactured using 3D printing and laser cutting. The setup is placed on a structure built with T-slotted aluminium profiles that allow the proper levelling of the channel, as shown in Fig.~\ref{fig:angular-vertical-mov}. 

The soft elastomer base that rests over the acrylic base, as can be seen in Figs.~\ref{Fig:01}(b) and ~\ref{Fig:01}(d), is made of a polyaddition polymer made of room-temperature polymerising silicone from Esprit Composite (Shore hardness scale OO). This material can reproduce fine details through moulding. It is also resistant to ageing and moisture. The base of the synchrowave comprises two 3-mm-wide ring-shaped acrylic layers, on which the pistons of the servomotors fit in such a way that popper motion is guaranteed. 

We used 64 coreless Dynamixel AX -- 18A servomotors, manufactured by \textsc{Robotis}, connected in series and fixed to the acrylic base channel by rubber restraints, and fed by eight power supplies of 12V and 29A regularly distributed along the loop. Each servomotor can process one bit of data sequentially through a single communication channel composed of a receptor and a transmitter. Each servomotor has a range of movement between 0 and 300$^\circ$, with an approximate resolution of 0.29$^\circ$. Each unit of the structure independently converts the servomotor's orbital motion to the piston's vertical motion. The units contain steel shafts, laser-cut acrylic and 3D-printed PLA/ABS parts. 

Servomotors are controlled through the microcontroller OpenCM 9.04 using an open-source code written in Arduino IDE. This code assigns an ID number $i$ between 0 and 63 to each servomotor according to its position along the synchrowave. The servomotors' rotational motion is responsible for generating the pistons' vertical displacement. The vertical position of $i$-th motor as a function of time and space is defined as
\begin{equation}
    z_i = A\sin\left( 2\pi f \dfrac{t}{1000} - \dfrac{2\pi}{\lambda_n}i\right),
\end{equation}
where $A$ is the bottom movement amplitude in millimetres, $f$ is the rotation frequency in Hertz, $t$ is time in milliseconds, $i$ is the servomotor ID, and $\lambda_n$ is the number of servomotors that complete one spatial period. To satisfy the periodic boundary conditions, $\lambda_n$ must be a power of 2 with $2 \leq \lambda_n \leq64$. If $\lambda_n = 2$, then 32 wavelengths fit in the synchrowave, while for $\lambda_n = 64$, one single long wavelength fits in the whole domain.

Each servomotor receives the maximum angular position as input, which can be mapped to a unique piston vertical position. These parameters are related after the system geometry shown in Fig.~\ref{fig:angular-vertical-mov}: When the motor is positioned in its initial position, the connecting rod has a 45$^\circ$ angle with the horizontal. Then, the angular position of the $i$-th motor is related to its piston position $z_i$ by
\begin{equation}
    \theta_i = \arccos \left( \dfrac{z_i^2+r^2-b^2}{2rz_i} \right)\dfrac{180^\circ}{\pi}+45^\circ.
\end{equation}
The code we implemented allows real-time control of the frequency, wavelength, and amplitude. The amplitude is limited by the dimensions of the system of motion transfer, and the maximal frequency is limited by the servomotors' response time. There is also a constraint for the maximal torque, but all the measurements are below this limit.

\begin{figure}
    \centering
\includegraphics[width=\linewidth]{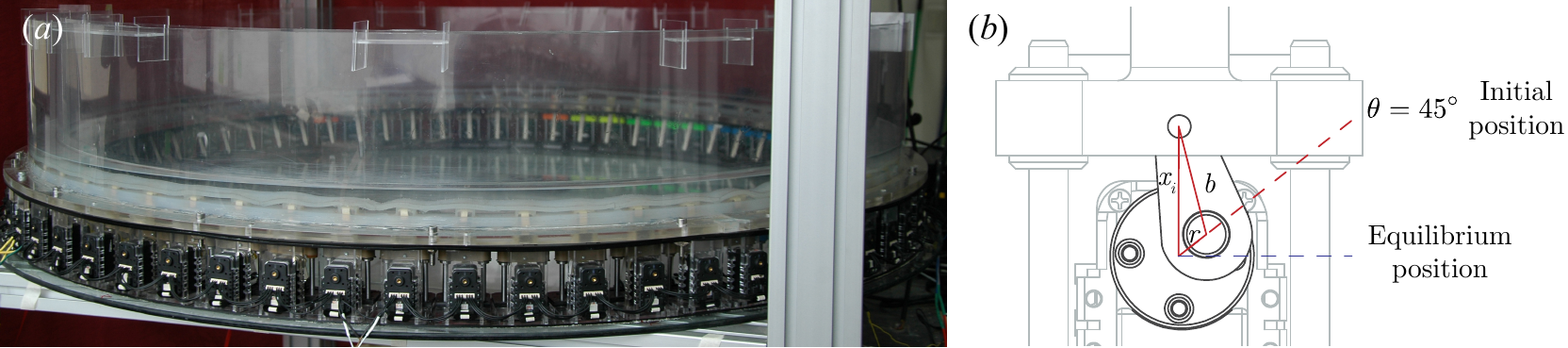}
    \caption{\textbf{Experimental development of the synchrowave}: \textbf{(a)} Photograph of the synchrowave tabletop setup. \textbf{(b)} Transmission system to transform angular motion to vertical at each building block. The pivoting point is $r=5$-mm apart from the centre of rotation of the servomotor, and the connecting rod is $b=17-$mm long.}
    \label{fig:angular-vertical-mov}
\end{figure}

To characterise the experimental gain of water waves in the synchrowave, we recorded the motion of the free surface by locating a high-speed camera and a backlit LED panel in a section of the channel, as shown in Fig.~\ref{Fig:01}(a). With a wavelength of 49.09~cm and 1.3~cm water height, we performed a frequency sweep between 0.5 and 1.3 Hz, recording 1024$\times$520-pixel-resolution videos for each frequency (1 pixel corresponds to 0.0624 mm). The sampling frequency is adjusted to fully resolve the bottom motion by recording 256 frames in 4 periods. In each video frame, we used an image detection algorithm to find the position of the free surface and bottom. The data is conveniently plotted in spatiotemporal diagrams for each frequency, showing the free surface motion as a function of time and space, as shown in Fig.~\ref{Fig:02}(\textit{a}). In addition to the bottom displacement data, this is used to derive the experimental amplitudes, $\eta_0$ and $\zeta$, defined as the peak-to-peak displacement of the wave and the bottom. The experimental gain $G$ for each frequency is simply defined as
\begin{equation}
    G(\omega) = \dfrac{\eta_0}{\zeta}.
\end{equation}
The experimental phase can also be obtained straightforwardly by measuring the time shift between the instants at which the wave and bottom displacement reach their maximal value throughout a cycle.

\section*{Calculation of the gain of the synchrowave}

The synchrowave fluid mechanics is governed by the Navier-Stokes (NS) equations for incompressible fluids,
\begin{subequations}
\label{eq:Ns-incompress}
\begin{equation}
\label{eq:NS}
\partial_{t}\mathbf{u}+(\mathbf{u}\cdot\boldsymbol{\nabla})\mathbf{u}=-\frac{1}{\rho}\boldsymbol{\nabla} P+\nu\nabla^{2}\mathbf{u}+\mathbf{g},
\end{equation}
\vspace{-0.7 cm}
\begin{equation}
\label{eq:incompress}
\boldsymbol{\nabla}\cdot\mathbf{u}=0,
\end{equation}
\end{subequations}
where $\mathbf{u}=u\mathbf{i}+w\mathbf{k}$ is the velocity field, $P$ is the pressure, $\nu$ is the kinematic viscosity, $\mathbf{g}=-g\mathbf{k}$ is the acceleration of gravity, and $\rho$ is the density of the fluid (water). We shall assume that the system is invariant along the $y$-direction, i.e. perpendicular to the walls.

The input drive of the system is assumed to follow the pedal-wavemaker driving\cite{Vivanco2021}, which assumes an orbital periodic motion at each point of the water bed with a phase that travels at a constant speed. This approximation for our synchrotron will be justified later. Thus, the vertical motion of the bottom is given by equation~\eqref{Eq:InputVertical} satisfying the no-slip boundary condition at the bottom,
\begin{equation}
	\label{eq:noslip2}
	\partial_t z_b = \left.w\,\right|_{z=-h},
\end{equation}
whereas the horizontal periodic displacements of pedal wavemakers, which can be described by the complex field $x_b$, is given by
\begin{equation}
x_b(x,t)=\mathrm{i}\chi\exp[\mathrm{i}\left(kx-\omega t\right)],\label{Eq:InputHorizontal}
\end{equation}
where $\chi$ is the amplitude of the horizontal component of the pedalling motion. The horizontal input drive must also comply with the no-slip boundary condition at the bottom,
\begin{equation}
	\label{eq:noslip1}
	\partial_tx_b=\left.u\,\right|_{z=-h}.
\end{equation}
The linear response under study, i.e. the output of the system of equations~\eqref{eq:Ns-incompress}, to the input drive of equation~\eqref{Eq:InputVertical} is a surface gravity wave, denoted as $\eta$, which in the linear regime has the same frequency $\omega$ and wavenumber $k$ as the bottom motion,
\begin{equation}
\eta(x,t)=\eta_0\exp[\mathrm{i}\left(kx-\omega t\right)],\label{Eq:Output}
\end{equation}
where $\eta_0$ is a complex amplitude. At the top of the fluid domain, the velocity field must comply with linearised boundary conditions for a free surface, namely
\begin{subequations}
	  	\label{eq:boundary}
	\begin{equation}
		\label{eq:kinematic}
		\partial_t\eta=\left.w\,\right|_{z=\eta},
	\end{equation}
 \vspace{-0.5 cm}
	\begin{equation}
		\label{eq:normal_stress}
		\left.2\nu\partial_{z}w-\frac{P}{\rho}\,\right|_{z=\eta}=0,
	\end{equation}
 \vspace{-0.5 cm}
\begin{equation}\label{eq:tangential_stress}\left.\partial_{x}w+\partial_{z}u\,\right|_{z=\eta}=0.
	\end{equation}
\end{subequations}
where equation~\eqref{eq:kinematic} follows from the kinematic boundary condition, and equation~\eqref{eq:normal_stress} and \eqref{eq:tangential_stress} from normal and tangential stress balance. Following Vivanco et al.\cite{Vivanco2021}, we solve the linearised version of equations~\eqref{eq:Ns-incompress} using the Helmholtz decomposition \cite{Lamb1932},
\begin{equation}
\label{Eq:Helmholtz}
\ensuremath{\mathbf{u}=\boldsymbol{\nabla}\phi+\boldsymbol{\nabla}\times\left(\psi\mathbf{j}\right)},
\end{equation}
where $\phi$ is the velocity potential and $\psi$, the Stokes stream function. After replacing the Helmholtz decomposition in the governing equations, the velocity potential $\phi$ must satisfy,
\begin{subequations}
\label{Eq:Potential}
    \begin{equation}
	\label{eq:incompress2}
	    \nabla^2\phi =0,
	\end{equation}
  \vspace{-0.5 cm}
    \begin{equation}
		\label{eq:potential}
		\partial_t\phi+\frac{P}{\rho}+gz=0,
	\end{equation}
\end{subequations}
while the diffusion equation governs the stream function $\psi$,
\begin{equation}
    \label{eq:stream}
	\partial_t\psi=\nu\nabla^2\psi.
\end{equation}
Accordingly, the exact solutions of the velocity potential and the stream function are of the form \cite{Vivanco2021},
\begin{subequations}
\label{Eq:FieldSolutions}
\begin{equation}
    \phi(x,z,t)= \frac{\omega}{k} \left[ A\cosh(kz) + B\sinh(k z)\right]\exp[\mathrm{i}(k x - \omega t)],\label{eq:phi_xy}
\end{equation}
 \vspace{-0.5 cm}
\begin{equation}
    \psi(x,z,t)= \frac{\omega}{k} \left[ C\cosh(m z) + D\sinh( m z)\right]\exp[\mathrm{i}(k x - \omega t)],\label{eq_psi_xy}
\end{equation}
\end{subequations}
where $m =\sqrt{k^2 - \mathrm{i}\omega/\nu}$ is a complex wavenumber that accounts for viscous effects.

Hereon, we use the dimensionless wavenumber $\overline{k}=kh$ and $\overline{m}=mh$, and the dimensionless frequency $\overline{\omega}=\omega/\omega_0$, where $\omega_0=\sqrt{gk\tanh{kh}}$ is the dispersion relation of gravity waves. Likewise, the amplitude of the surface wave, the horizontal bottom motion, the vertical bottom motion, the vertical coordinate, and the horizontal coordinate in dimensionless form are $\overline\eta_0=\eta_0/h$,  $\overline\chi=\chi/h$,  $\overline\zeta=\zeta/h$, $\overline{z}=z/h$, and $\overline{x}=x/h$, respectively.

Replacing the solutions of equation~\eqref{Eq:FieldSolutions} the boundary conditions of equations~\eqref{eq:boundary}, we obtain the following linear system of equations for the dimensionless vector $(\overline A,\,\overline B,\,\overline C,\,\overline D)^{T}\equiv( A,\, B,\,C,\, D)^{T}/h$,
\begin{equation}
\begin{pmatrix}\overline\omega\overline\Theta\beta\overline{k}^{3/2} & i & 1 & -i\alpha\overline{k}^{3/2}\overline\Theta\\
0 & i & \beta & 0\\
i\cosh\overline k & -i\sinh\overline k & -\alpha\sinh\overline m & \alpha\cosh\overline m\\
-i\sinh\overline k & i\cosh\overline k & \cosh\overline m & -\sinh\overline m
\end{pmatrix}\begin{pmatrix}\overline A\\
\overline B\\
\overline C\\
\overline D
\end{pmatrix}=\begin{pmatrix}0\\
0\\
\overline\chi\\
\overline\zeta
\end{pmatrix},\label{eq:LinearSystem}
\end{equation}
with $\alpha\equiv m/k$, $\beta\equiv1-i/\overline{\theta}$, and $\overline \Theta \equiv 2\nu/\sqrt{gh^3}$ is the dimensionless viscosity of the fluid. After substitution of the solution to equation~\eqref{eq:LinearSystem} into the kinematic boundary condition \eqref{eq:kinematic}, one obtains
\begin{subequations}
\label{Eq:Constants}
\begin{align}
    A&= \frac{i\alpha\overline\theta}{\Gamma_B^{(0)}(\overline\theta -i)}\left(\zeta\cosh\overline{k}+\frac{\left(\alpha\overline\theta^2\overline\omega^2\Gamma_B^{(1)}-\Gamma_B^{(0)}-\Gamma_S^{(0)}\right)\mathcal{P}}{\alpha\overline\theta^2\overline\omega^2(\Gamma_S+\overline\theta\Gamma_B)}\right),\label{eq:A}\\
    B&=\frac{(1+i\overline\theta)\mathcal{P}}{\overline\theta(\Gamma_S+\overline\theta\Gamma_B)},\label{eq:B}\\
    C&=D=\frac{\mathcal{P}}{\Gamma_S+\overline\theta\cosh\overline m\left[\alpha\sinh\overline k + (1+\alpha\sigma)\cosh\overline k\right]-\overline\theta\sinh\overline m\left[\cosh\overline k+(\alpha+\sigma)\sinh\overline k\right]},\label{eq:CandD}
\end{align}
\end{subequations}
with $\overline\theta=\overline \Theta \overline k^{3/2}/\overline\omega$, 
 $\sigma=\overline\omega^2(\overline\theta-i)^2$, and $\mathcal{P}=\overline\zeta[\overline\theta\cosh\overline{k}-(\overline\theta -i)\cosh\overline m]$. The constants $\Gamma_S^{(0)}$ and $\Gamma_S$ are related to the angular frequency, depth, wavenumbers and viscosity through $\Gamma_S^{(0)}=(i-\overline\theta)\alpha\overline\theta\overline\omega^2$ and 
$\Gamma_S=(i-\overline\theta)\left(1+\alpha\overline\theta^2\overline\omega^2\right)$, whereas the remaining constants are given by $\Gamma_B^{(0)}=\alpha\sinh\overline k\cosh\overline m-\cosh\overline k\sinh\overline m$, $\Gamma_B^{(1)}=\alpha\sinh\overline k\sinh\overline m-\cosh\overline k\cosh\overline m$, and  $\Gamma_B=\Gamma_B^{(0)}-\Gamma_B^{(1)}+\sigma(\alpha\cosh\overline k \cosh\overline m-\sinh\overline k\sinh\overline m)$.

The velocity potential and the stream function can be built by replacing the expressions in equation~\eqref{Eq:Constants} into equation~\eqref{Eq:FieldSolutions}. The complete velocity field can be computed through equation~\eqref{Eq:Helmholtz}. Finally, the kinematic boundary condition \eqref{eq:kinematic} implies that $\overline\eta_0=C+iB$, which can be written as $\overline\eta_{0}=G_x\overline\chi+G_z\overline\zeta$, where $G_x$ and $G_z$ are the gain due to the horizontal and vertical component of the pedal-wavemaker motion, respectively. Finally, one obtains
\begin{subequations}
\begin{align}
 G_x(\overline\omega, \overline k)&=\frac{\mathrm{i}\overline\Theta}{\Xi(\overline\omega, \overline k)}\left(\frac{1}{\alpha}\sinh\alpha \overline{k}-\sinh \overline{k}\right),\\
 G_z(\overline\omega, \overline k)&=\frac{\left(1+\mathrm{i}\overline\Theta\right)\cosh\alpha\overline{k}-\mathrm{i}\overline\Theta\cosh\overline{k}}{\Xi(\overline\omega, \overline k)},
\end{align}
\end{subequations}
with the function $\Xi(\overline\omega, \overline k)$ defined as
\begin{equation}
\Xi(\overline\omega, \overline k)\equiv\cosh\alpha\overline{k}\cosh\overline{k}\left(1-\frac{1}{\overline\omega^{2}}\right)\left(1-\frac{\mbox{tanhc}\,\alpha \overline{k}}{\mbox{tanhc}\,\overline{k}}\right)+\frac{\mbox{sinhc}\,\alpha \overline{k}}{\mbox{sinhc}\,\overline{k}}
+\overline{k}^{2}\left[\mbox{sinhc}^2\left(\frac{\overline{k}(\alpha +1)}{2}\right)+\mbox{sinhc}^2\left(\frac{\overline{k}(\alpha -1)}{2}\right)\right].
\end{equation}
A graphical inspection of the plots of the expressions above for a fixed wavelength shows that $ G_z/G_x\gg1$ for all $\overline\omega$, and thus, the vertical motion of the pedal wavemakers is the main contributor of the total gain by far. Therefore, the theory of pedal wavemakers provides a good framework for modelling the vertical input drive of the synchrotron servomotors. Accordingly, the total gain $G$ is well approximated by $G_z$, and thus by the entire expression in equation~\eqref{YAmplitude}.

\section*{Curvature effects on gravity waves}
\label{CurvatureTheory}

To get an insight into the effects of the wall curvature on the dynamics of gravity waves in our synchrotron, we study the three-dimensional free-surface equations in cylindrical coordinates. Let $R_1$ and $R_2$ be the radius of the inner and outer walls of the synchrotron, respectively.  In Cartesian coordinates, the dynamic and kinematic boundary condition now reads as follows,
\begin{subequations}
\label{eq:surface0}
\begin{align}
\left.\partial_{t}\phi+\frac{1}{2}\left[\left(\partial_{x}\phi\right)^{2}+\left(\partial_{y}\phi\right)^{2}+\left(\partial_{z}\phi\right)^{2}\right]+g\eta\right|_{z=\eta} & = 0,\label{eq:dynamic0}\\
\left.\partial_{t}\eta+\partial_{x}\phi\partial_{x}\eta+\partial_{y}\phi\partial_{x}\eta-\partial_{z}\phi\right|_{z=\eta} & = 0.\label{eq:kinematic0}
\end{align}
\end{subequations}
For simplicity, we omit the bottom motion in this calculation since we are interested in the curvature effects on the surface waves via their dispersion relation. Thus, at the bottom of the channel, we have
\begin{equation}
\label{eq:bottom}
\left. \partial_{z}\phi \right|_{z=-h} = 0,
\end{equation}
whereas at the walls of the synchrotron, we have
\begin{subequations}
\label{eq:walls}
\begin{align}
\left.\left.x\partial_{x}\phi+y\partial_{y}\phi\right|\right|_{x^{2}+y^{2}=R_{1}^{2}} &= 0.\label{eq:wall1}\\
\left.\left.x\partial_{x}\phi+y\partial_{y}\phi\right|\right|_{x^{2}+y^{2}=R_{2}^{2}} &= 0.
\label{eq:wall2}
\end{align}
\end{subequations}
Equations~\eqref{eq:surface0} and the incompressibility condition \eqref{eq:incompress2} written in the cylindrical coordinates $(r,\varphi, z)$ reads as follows,
\begin{subequations}
\begin{align}
    \left.\partial_{t}\phi+\frac{1}{2}\left[\left(\partial_{r}\phi\right)^{2}+\left(\frac{1}{r}\partial_{\varphi}\phi\right)^{2}+\left(\partial_{z}\phi\right)^{2}\right]+g\eta\right|_{z=\eta} & = 0,\label{eq:dynamic0-2}\\
\left.\partial_{t}\eta+\partial_{r}\phi\partial_{r}\eta+\frac{1}{r^2}\partial_{\varphi}\phi\partial_{\varphi}\eta-\partial_{z}\phi\right|_{z=\eta} & = 0,\label{eq:kinematic0-2}\\
\frac{1}{r}\partial_{r}\left(r\partial_{r}\phi\right)+\frac{1}{r^{2}}\partial_{\varphi\varphi}\phi+\partial_{zz}\phi & = 0,\label{eq:laplace-2}    
\end{align}
\end{subequations}
whereas the boundary conditions at the curved walls, equations~\eqref{eq:walls}, gives $\partial_{r}\phi=0$ at $r=R_{1}$ and $r=R_{2}$. Linearising equations~\eqref{eq:dynamic0-2} and~\eqref{eq:kinematic0-2}, we obtain
\begin{subequations}
    \begin{align}
        \left.\partial_{t}\phi+g\eta\right|_{z=0} & = 0,\label{eq:dynamic0_lin}\\
        \left.\partial_{t}\eta-\partial_{z}\phi\right|_{z=0} & = 0.\label{eq:kinematic0_lin}
    \end{align}
\end{subequations}
Differentiating equation~\eqref{eq:dynamic0_lin} with respect to time, we find upon substitution of equation~\eqref{eq:kinematic0_lin} 
\begin{equation}
\left.\partial_{tt}\phi+g\partial_{z}\phi \right|_{z=0}  =0,
\end{equation}
with solutions of the form
\begin{equation}
\phi\left(r,\varphi,z,t\right)=\frac{\cosh\left[k\left(z+h\right)\right]}{\cosh kh}\exp\left[i\left(m\varphi-\omega t\right)\right]R\left(r\right),
\end{equation}
where $\omega^{2}=gk\tanh kh$ is the classical water-wave dispersion relation. On the other hand, the function $R(r)$ is given by the solutions of the Bessel differential equation,
\begin{equation}
\label{Eq:Bessel}
    \xi^{2}\frac{\hbox{d}^2f}{\hbox{d}\xi^2}+\xi\frac{\hbox{d}f}{\hbox{d}\xi}+\left(\xi^{2}-m^{2}\right)f=0,
\end{equation}
where $\xi\equiv kr$ is a dimensionless variable such that  $R(r)=f(\xi)$. Thus, $R\left(r\right)=\alpha_{m}J_{m}\left(kr\right)+\beta_{m}Y_{m}\left(kr\right)$, where $J_{m}(\xi)$ and $Y_{m}(\xi)$ are the Bessel functions of the first and second kind, respectively. The boundary conditions at the curved walls require 
\begin{equation}
k\left(\alpha_{m}\frac{\hbox{d}J_{m}}{\hbox{d}\xi}+\beta_{m}\frac{\hbox{d}Y_{m}}{{\hbox{d}\xi}}\right)=0,\quad\text{ for }\quad\xi=kR_{1}\text{ and }\xi=kR_{2}.
\end{equation}
This condition is met if
\begin{equation}
\label{eq:DiffRelation}
    \left.\frac{\hbox{d} Y_{m}}{\hbox{d}J_{m}}\right|_{\xi=kR_{1}}^{\xi=kR_{2}}=0.
\end{equation}
Defining the mean radius of the synchrotron and the water-channel half-width of the  as $\overline{R}=\left(R_{1}+R_{2}\right)/2$ and $e=\left(R_{2}-R_{1}\right)/2$, respectively, and performing the Taylor expansion of equation~\eqref{eq:DiffRelation} for $\lambda\gg e$ (i.e., $ke\sim0$), we obtain
\begin{equation}
\left.ke\frac{\hbox{d}}{\hbox{d}\xi}\left(\frac{\hbox{d}Y_{m}}{\hbox{d}J_{m}}\right)\right|_{\xi=k\overline{R}}+\mathcal{O}\left(k^{2}e^{2}\right) =0.
\end{equation}
Developing the derivatives and using equation~\eqref{Eq:Bessel} leads to
\begin{equation}
    \left.\frac{2}{\pi\xi}\left(1-\frac{m^{2}}{\xi^{2}}\right)\left(\frac{\mbox{d}J_{m}}{\mbox{d}\xi}\right)^{-2}\right|_{\xi=k\overline{R}}+\mathcal{O}\left(k^{2}e^{2}\right) =0.
\end{equation}
Thus, we have $k\overline{R}=m+(1/3!)(\left(ke\right)^{2})/m+\mathcal{O}\left(k^{4}e^{4}\right)=0$, from which follows that
\begin{equation}
    \frac{k\overline R}{m}=\left(1+\frac{1}{6}\frac{e^{2}}{\overline{R}^{2}}+...\right).
\end{equation}
The later expression shows that in an annular waveguide, the classical water-wave dispersion relation $\omega^{2}=gk\tanh kh$ remains valid, but for a modified wavenumber $k$, that at leading order is the azimuthal wavenumber multiplied by the waveguide mean radius, $m\overline R$. Notice that this value corresponds to the perimetral wavenumber (the wavelength along the arclength is $\lambda_p =2\pi\overline R / m$). The correction terms are powers of the channel width-to-diameter ratio, and the leading one is of quadratic order. For our setup, the corrections are of the order of $10^{-4}$ or higher, and thus curvature effects are negligible. 

\section*{Numerical SPH formulation}
\label{SPHFormulation}

Our numerical setup for SPH simulations is schematised in Fig.~\ref{Fig:05}(\textit{a}). A fluid layer of density $\rho_0=1\,\hbox{gr/cm}^3$ and depth $h=1\,\hbox{cm}$ is initially at rest in a fluid domain of length $L=20\,\hbox{cm}$, that neglects curvature effects.
In the SPH method, the density, pressure, and velocity field variables and gradients are obtained numerically from a Lagrangian formulation of the Navier-Stokes equations using a combination of two approximation techniques: the \emph{kernel approximation} and the \emph{particle approximation}. SPH provides a powerful method for the simulation of fluids under high deformations, such as free surface waves, shock waves, high vorticity flows, and spraying, all very common phenomenons in the study of coastal and ocean dynamics \cite{Sigalotti2006, Sigalotti2009, Sun2018, Gnanasekaran2019}. The SPH method can be highly dissipative without modern amending techniques compared to other computational fluid dynamics methods \cite{McCue2006}. This combination of features poses SPH as an appropriate framework for testing the robustness of the resonance phenomena in our synchrowave regarding viscosity. Here, we briefly overview the SPH formulation used in this work, which was based on the Open Software PySPH \cite{PySPH}. For further details, Liu and Liu give a comprehensive and detailed presentation of the SPH method in Ref.~\cite{Liu2010}, and Monaghan in Refs.~\cite{Monaghan2005, Monaghan2012}.

\begin{figure*}
    \centering
    \includegraphics[width=\linewidth]{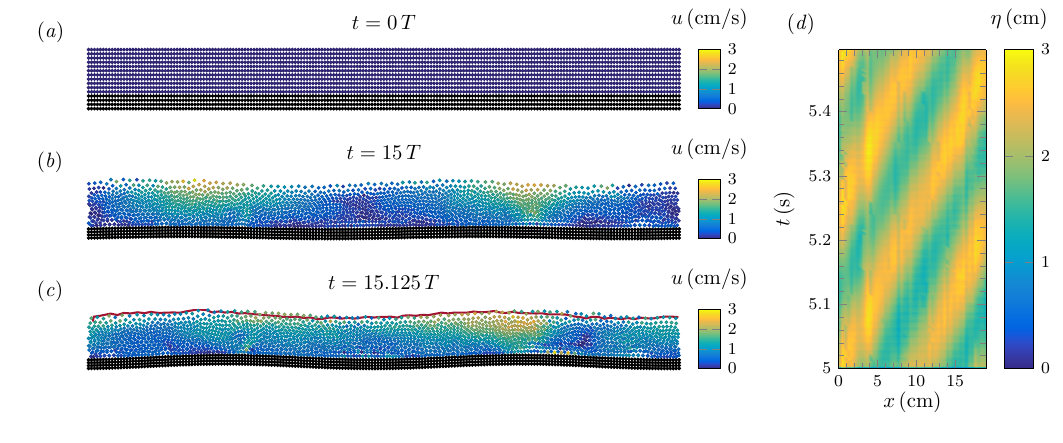}
    \caption{\textbf{Numerical protocol for simulating the synchrotron water waveguide}. \textbf{(\textit{a})} Numerical setup used in SPH simulations. Waves are generated by the vertical motion of the bed (black solid particles) in a two-dimensional fluid layer (turquoise fluid particles) of depth $\overline{h}=1$ and length $\overline{L}=20$. \textbf{(\textit{b})} The numerical simulation begins with a gradual increase in the amplitude of the bottom motion. At some $t>0$, one observes a regular travelling wave at the free surface. \textbf{(\textit{c})} Surface detection algorithm performed on a fully developed gravity wave with $\overline{\lambda}=10$. \textbf{(\textit{d})} Spatiotemporal diagram obtained from the concatenation of consecutively detected free surfaces. The depicted case corresponds to $\overline{\omega}=1.01$, i.e. close to resonance.}
    \label{Fig:05}
\end{figure*}

The SPH method for numerically solving the hydrodynamic equations is based on two main approximations: the kernel approximation and the particle approximation.
In the kernel approximation, the continuity and momentum equations are multiplied by a kernel function $W=W(|\mathbf{r}-\mathbf{r}'|;h_w)$ and integrated along $\mathbf{r}'$ in the fluid domain, leading to the so-called weighted integrals\cite{Liu2010}. The kernel $W$ is a smooth, normalised, and monotonically decaying function of the distance $|\mathbf{r}-\mathbf{r}'|$, typically with a bell-like shape. Formally, it is assumed that $W(|\mathbf{r}-\mathbf{r}'|;h_w)\to\delta(|\mathbf{r}-\mathbf{r}'|)$ as $h_w\to0$, where $\delta$ is the Dirac delta function\cite{Monaghan2012}. Moreover, kernels are usually chosen to vanish for $|\mathbf{r}-\mathbf{r}'|>h_w$, i.e. they have a \emph{compact support domain} of radius $h_w$ around $\mathbf{r}'$. This first approximation step gives the field variables and gradients as smoothed, local space-averaged values around each fluid particle placed at $\mathbf{r}'$, where only close-enough fluid particles contribute to the space averaging. The radius $h_w$ of the compact support domain is known as the \emph{smoothing length}. We used a cubic-spline kernel with a smoothing length of $h_w=0.13\,\hbox{cm}$ in our SPH simulations.

The formal space discretisation of the fluid domain is performed using particle approximation, and weighted integrals are estimated as discrete sums on a set of fluid particles. We used $N=2343$ fluid particles and $852$ solid particles simulating the soft elastomer bottom, all with diameter $D=0.1\,\hbox{cm}$. In this pure SPH formulation, the soft elastic bottom is represented by discrete solid particles. We use three layers of solid SPH particles for better performance of the bottom boundary and to avoid the percolation of fluid particles during the bottom motion. The boundary condition at the moving bottom is formulated in terms of a generalised wall boundary condition   \cite{Adami2012}, which is based on a local force balance between the bottom solid particles and the fluid particles to prevent wall penetration. We implemented the weakly-compressible SPH formulation   \cite{Hughes2010} for better computational performance in the simulation of the free-surface flow, keeping the density fluctuations small ($\sim 1\%$) \cite{Monaghan2005}. The resulting discrete SPH formulation for the continuity and momentum equations for the $i$-th particle is
\begin{subequations}
\label{Eq:SPHEquations}
\begin{align}
\label{Eq:SPHContinuity}
    \frac{\mathrm{d}\rho_i}{\mathrm{d}t}&=\sum_{j\neq i}m_j\mathbf{u}_{ij}\cdot\boldsymbol{\nabla}_iW_{ij}(h_w),\\
\label{Eq:SPHMomentum}
    \frac{\mathrm{d}\mathbf{u}_i}{\mathrm{d}t}&=-\sum_{j\neq i}m_j\left(\frac{P_i}{\rho_i^2}+\frac{P_j}{\rho_j^2}\right)\boldsymbol{\nabla}_iW_{ij}(h_w)+\boldsymbol{\Pi}_i+\mathbf{g}_i,
\end{align}
\end{subequations}
where $W_{ij}(h_w)=W(\mathbf{r}_{ij};h_w)$, $\mathbf{r}_{ij}=\mathbf{r}_i-\mathbf{r}_j$, $\mathbf{u}_{ij}=\mathbf{u}_i-\mathbf{u}_j$, and $\mathbf{\nabla}_i$ denotes a gradient with respect to the coordinates of the $i$-th particle. The artificial viscous acceleration $\Pi_{i}$ is a numerical term originally introduced to control nonphysical instabilities appearing in problems with shock waves \cite{Monaghan2005, Liu2010, Monaghan2012, PySPH}. We use the SPH formulation of the entropically damped artificial compressibility term \cite{Ramachandran2019, PySPH},
\begin{equation}
\label{Eq:ViscositySPH}
\boldsymbol{\Pi}_{i}:=2\alpha h_w C\sum_{j\neq i}\frac{m_j}{\rho_i+\rho_j}\,\frac{\mathbf{r}_{ij}\cdot\boldsymbol{\nabla}_iW_{ij}(h_w)}{\left(|\mathbf{r}_{ij}|^2+\epsilon h_w^2\right)}\mathbf{u}_{ij},
\end{equation}
where $\alpha=0.2$, $C=442.94\,\mbox{cm/s}$, and $\epsilon=0.01$ is a parameter introduced to avoid singularities for vanishing $|\mathbf{r}_{ij}|$. The pressure at the $i$-th particle is specified by its density via the usual equation of state for weakly compressible fluids, namely
\begin{equation}
    \label{Eq:EqState}
    P_i=\frac{P_0}{\gamma}\left[\left(\frac{\rho_i}{\rho_0}\right)^{\gamma}-1\right],
\end{equation}
where $P_0$ and $\rho_0$ are the reference pressure and density, respectively. For fluids, it is common to use $\gamma=7$. Finally, time integration of equations~\eqref{Eq:SPHEquations} is achieved using standard time-integration methods for ordinary differential equations. This work used a second-order predictor-corrector integrator with time-step $\Delta t=1\times10^{-5}\,\hbox{s}$. We have chosen $P_0$ in equation~\eqref{Eq:EqState} large enough to keep the density fluctuations around $1\%$ \cite{Monaghan2005}.

Every numerical simulation starts with a gradual increase of the amplitude of the vertical oscillations at the bottom from zero to a fixed maximum value, $\overline{A}=A/h=0.042$. As this happens, the system responds with a sustained build-up of a smooth surface wave without neither defects or fronts travelling throughout the numerical domain, as depicted in Fig.~\ref{Fig:05}(\textit{b}). Once gravity waves form, the surface profiles are first extracted from a boundary-detection algorithm applied to the set of SPH fluid particles [see Fig.~\ref{Fig:05}(\textit{c})]. The boundary detection is performed at time steps given by $\Delta\overline{T}=\overline{T}/8$, where $\overline{T}$ is the period of the bottom motion. Our measurements account for an integer number of oscillations of the servomotors. A typical spatiotemporal diagram of the waves as they travel streamwise is depicted in Fig.~\ref{Fig:05}(\textit{d}). As the output of SPH numerical is Lagrangian, i.e. a set of particles located throughout the domain, each carrying its hydrodynamic quantities, the measurement of the wave response requires post-processing. After applying a boundary-detection algorithm to the set of SPH fluid particles to detect the free surface at each time step, we generate a spatiotemporal diagram collecting all the detected boundaries in time, as shown in Fig.~\ref{Fig:05}(\textit{d}). Then, we compute the two-dimensional Fast Fourier Transform in space and time of such spatiotemporal diagram and find its maximum, which we denote as $\mathcal{F}_{\max}(\eta_{envl})$. Following equation~\eqref{Eq:InputVertical}, the bottom moves as $\eta_{b}(x,t)=A\cos(kx-\omega t)\propto \mbox{Re}{\,[z_b(x,t)]}$, i.e., a wave force travelling towards the streamwise direction. We compute similarly the maximum of the fast Fourier transform of this pre-programmed spatiotemporal dynamics corresponding to the servomotors, $\mathcal{F}_{\max}(\eta_{b})$. Finally, we obtain the gain and the phase of the system with $G=\left\vert \mathcal{F}_{\max}(\eta_{envl})\right\vert/\left\vert\mathcal{F}_{\max}(\eta_{b}) \right\vert$ and $\phi=-\arg\left[\mathcal{F}_{\max}(\eta_{envl})/\mathcal{F}_{\max}(\eta_{b})\right]+\Delta\phi_0$, respectively, where $\Delta\phi_0$ is the accumulated phase shift from the beginning of the simulation to the first boundary detection.
}

\section*{Acknowledgments}
The authors thank Enrique Cerda, Claudio Falcón, Christophe Josserand and Thomas Séon for fruitful discussions on the experimental setup and its reach. The authors are also grateful to Bas Dijkhuis,  Narek Halsdorfer, Jeremy Riffo, Lucas Martínez, Giuliana Álvarez, for their technical help in building the first synchrowave. J.F.M. and A.E. acknowledge the financial support of Agencia Nacional de Investigación y Desarrollo (ANID—Chile) through the grant FONDECYT Postdoctorado No. 3200499. B.C. thanks the Australian Research Council Linkage Project Number LP190101283. L.G. and I.V. thank the FONDECYT/Iniciación grant No. 11170700 and No. 1221103.

\section*{Data Availability Statement}

The data that support the findings of this work are available from the corresponding author upon reasonable request.

\section*{Author contributions statement}

Conceptualisation, BC, LG and JFM; formal analysis, IV, JFM, and LG; experimental investigation, IV and LG; methodology, BC, JFM and LG; software, JFM and AE; supervision, JFM and LG; validation, JFM and LG; writing--original draft, JFM; writing--review and editing, IV and LG. All authors have read and agreed to the published version of the manuscript.

\section*{Competing interests} 

The authors declare no competing interests.

\providecommand{\noopsort}[1]{}\providecommand{\singleletter}[1]{#1}%
%



 


\end{document}